**SWSC**

 

# Space Weather impact on the degradation of NOAA POES MEPED proton detectors

Linn-Kristine Glesnes Ødegaard*, Hilde Nesse Tyssøy, Marit Irene Jakobsen Sandanger, Johan Stadsnes, and Finn Søraas

Birkeland Centre for Space Science (BCSS), Department of Physics and Technology, University of Bergen, Allégaten 55, N-5007 Bergen, Norway
*Corresponding author: linn-kristine.odegaard@uib.no

**ABSTRACT**

The Medium Energy Proton and Electron Detector (MEPED) on board the National Oceanic and Atmospheric Administration Polar Orbiting Environmental Satellites (NOAA POES) is known to degrade with time. In recent years a lot of effort has been put into calibrating the degraded proton detectors. We make use of previous work and show that the degradation of the detectors can be attributed to the radiation dose of each individual instrument. However, the effectiveness of the radiation in degrading the detector is modulated when it is weighted by the mean $ap$ index, increasing the degradation rate in periods with high geomagnetic activity, and decreasing it through periods of low activity. When taking $ap$ and the radiation dose into account, we find that the degradation rate is independent of spacecraft and detector pointing direction. We have developed a model to estimate the correction factor for all the MEPED detectors as a function of accumulated corrected flux and the $ap$ index. We apply the routine to NOAA POES spacecraft starting with NOAA-15, including the European satellites MetOp-02 and MetOp-01, and estimate correction factors.

## 1. Introduction

The space environment around the Earth is harsh, and instruments and satellites in orbit can be harmed if exposed to the strong radiation (see e.g. Horne et al. 2013). Here, we study the radiation damage experienced by the silicon solid state proton detectors, which are part of the Medium Energy Proton and Electron Detector (MEPED) on board the Polar Orbiting Environmental Satellites (POES) operated by the National Oceanic and Atmospheric Administration (NOAA).

NOAA POES are low Earth orbiting satellites with an approximate orbital altitude of 850 km. Fourteen spacecraft carry a designated detector suite to measure the flux of energetic charged particles at satellite altitude. The combined dataset from the MEPED detectors currently stretches over 37 years and is ideal for studies of long-term trends in space weather. However, Galand & Evans (2000) showed that the proton detector of the MEPED instrument suffered from degradation already after 1–2 years in operation. The degradation causes the instrument to seriously underestimate the incoming proton flux and makes the data unusable for quantitative studies. Lately, several studies have addressed the issue and estimated the degradation and how it evolves with time (Asikainen & Mursula 2011; Asikainen et al. 2012; Ødegaard 2013; Sandanger et al. 2015).

The goal of the studies by Asikainen & Mursula (2011), Asikainen et al. (2012) and Sandanger et al. (2015) is to extend the period of the quantitative applicability of the large MEPED dataset. The present study is driven by the same motivation, however, we aim to identify the controlling factors of the degradation. This is important because the detector correction from previous studies depends on new satellites being launched, and the last one is scheduled to launch in 2017.[1,2] We present a method to estimate the level of degradation of MEPED detectors which does not depend on the launch of new satellites.

In the following sections we will briefly present the NOAA POES satellites and the MEPED detectors. We will give some details on how the detectors degrade, and what has been done so far to correct the data. Finally, we will show how the degradation of the different energy channels of the detector can be related to the product of the accumulated corrected flux and the $ap$ index for the individual detectors. We present a method to calibrate the detectors on yearly basis.

## 2. NOAA POES and MEPED

The NOAA POES are meteorological satellites that provide global forecasts for NOAA's National Weather Service (NWS) with space-based Earth remote sensing. Today they provide data used for (among others) applications related to the oceans, detection of forest fires and monitoring the ozone hole in the Antarctica (Davis 2007).

In addition to the remote sensing instruments aimed for weather and climate data, a Space Environment Monitor (SEM) is included on the NOAA POES, which measures charged particles at satellite altitude. The purpose of this instrument is to reveal the impact of energetic charged particles on the upper atmosphere. The SEM instruments consist of several detectors, each designated to measure electrons or protons within a specific pointing direction and energy range. In this study we focus on the MEPED proton detector as this detector exhibits degradation over time. The MEPED electron detector is shielded by foil preventing protons with energy below 200 keV from entering and does not suffer from significant degradation. The SEM instrument package was updated after the launch of NOAA-14, and for MEPED this included changes in the energy channels and the viewing direction of the 90° telescope. To distinguish between the two versions satellites up to and including NOAA-14 are referred to as SEM-1 satellites, and SEM-2 starting with NOAA-15.

---

[1] http://www.eumetsat.int/website/home/Satellites/CurrentSatellites/Metop/index.html
[2] http://poes.gsfc.nasa.gov/

The MEPED instrument has two proton telescopes viewing almost perpendicular to each other. One telescope is pointing radially outward and is called the 0° detector. At high latitudes this telescope views particles near the centre of the atmospheric bounce loss cone. At high latitudes the second telescope, called the 90° detector, views particles at the edge and outside the atmospheric bounce loss cone. We focus mainly on the SEM-2 satellites, although the results will be applicable to the SEM-1 satellites as well. For the SEM-2 MEPED proton detector, measured particles are sorted in five differential energy channels and one integral energy channel, and the nominal energy thresholds of P1–P6 are listed in Table 1 (Evans & Greer 2000). The detector was designed to measure the intensity of protons ranging from 30 keV to greater than 6.9 MeV. With this wide energy range, both auroral protons, radiation belt protons and solar proton events can be measured by MEPED (detailed descriptions of the instruments can be found in Raben et al. (1995) for SEM-1 and Evans & Greer (2000) for SEM-2.

The operational lifetimes of all the MEPED instruments are plotted in Figure 1. MetOp-02 and MetOp-01 are meteorological polar orbiting satellites launched by the European Organisation for the Exploitation of Meteorological Satellites (EUMETSAT), but carry the SEM-2 suit from NOAA. The yearly sunspot number is plotted in Figure 1 for reference to illustrate how the combined dataset covers more than three solar cycles.

## 3. Data

We use the proton measurements provided by the NOAA's National Geophysical Data Center.[3,4] The proton data is given within the six energy channels presented in Table 1. The data are accumulated for 1 s, but the 0° and 90° detectors share electronics and a full dataset takes 2 s to obtain (Evans & Greer 2000). We calculate the monthly mean integral proton flux spectrum without the P6 channel, which is disregarded due to contamination from relativistic electrons (Yando et al. 2011). When not contaminated by electrons, P6 has very low count rates, and can be safely dropped in this context. The *ap* index is downloaded from the OMNI database, and the sunspot number is obtained from the SIDC.

## 4. Previous correction for degradation

The MEPED proton detector degrades in two known ways. A dead layer is created on the front detector, which increases with time. In addition, atoms making up the silicon detector itself are shifted in their place (Evans & Greer 2000; Galand & Evans 2000; McFadden et al. 2007).

---

[3] http://satdat.ngdc.noaa.gov/sem/poes/data/full/
[4] http://satdat.ngdc.noaa.gov/sem/poes/data/processed/ngdc/uncorrected/full/

**Table 1.** Energy thresholds of the MEPED proton detector (SEM-2).

| Channel | Nominal energy range of protons (keV) |
|---------|---------------------------------------|
| P1 | 30–80 |
| P2 | 80–250[a] |
| P3 | 250[a]–800 |
| P4 | 800–2500 |
| P5 | 2500–6900 |
| P6 | >6900[b] |

[a] We use 250 keV (Yando et al. 2011), as also used by Sandanger et al. (2015). The description of the instrument reports 240 keV (Evans & Greer 2000).
[b] According to simulations by Yando et al. (2011), P6 can measure protons with energy of at least 10 MeV. It is also sensitive to relativistic electrons.

When the dead layer of the detector thickens, more of the particle's energy is deposited in the detector without being measured. Defects in the atom structure deeper in the detector will result in a trapping of the charge carriers released in the detector in the form of electron-hole pairs, thus less of the incoming particle's energy is collected. Throughout most of the satellites' orbits the measured proton spectra are usually decreasing power-law. The degradation would then cause the flux measured between the nominal energy thresholds to be underestimated, as only higher energy particles are able to trigger the thresholds. In the South Atlantic Anomaly, where the proton differential spectra have increasing fluxes across P1 and P2, the degraded detectors may actually be overestimating the fluxes between the nominal energy thresholds. Overall, since the most common situation is a decreasing power-law proton spectrum, the effective energy thresholds of the instrument increase with time. The degradation is more severe in the lower channels (P1–P3) compared to the higher (P4–P5) because higher energy protons penetrate deeper into the detector where the damage is less prominent (Galand & Evans 2000; McFadden et al. 2007; Asikainen et al. 2012).

An aluminium coating on the detector stops protons with energy $E \lesssim 9$ keV from entering the detector (Seale & Bushnell 1987). Protons with $E > 9$ keV pass through the aluminium layer depositing some energy there. With this energy loss accounted for by setting the detector pulse height logic to 21.4 keV, the lower energy threshold of P1 is 30 keV (Evans & Greer 2000). The proton population with energies from 9 to 30 keV, which is stopped inside of the detector but not counted, is expected to be considerable. This low energy proton population might be of importance for the degradation, especially in the front where the dead layer is created (Galand & Evans 2000). It is not possible to use the MEPED data alone to get a reliable estimate of the particle flux with $E < 30$ keV, and the damaging effect of these fluxes cannot be directly accounted for.

Figure 2 illustrates how severe the degradation can be. Figure 2a shows daily mean proton fluxes in P1 (nominally 30–80 keV) for the entire operational period of NOAA-15, both detectors. NOAA-15 was launched in 1998 and is still actively measuring protons with MEPED, which makes it the only SEM-2 satellite active through more than one whole solar cycle (as can be seen in the sunspot number, plotted in Fig. 2c).

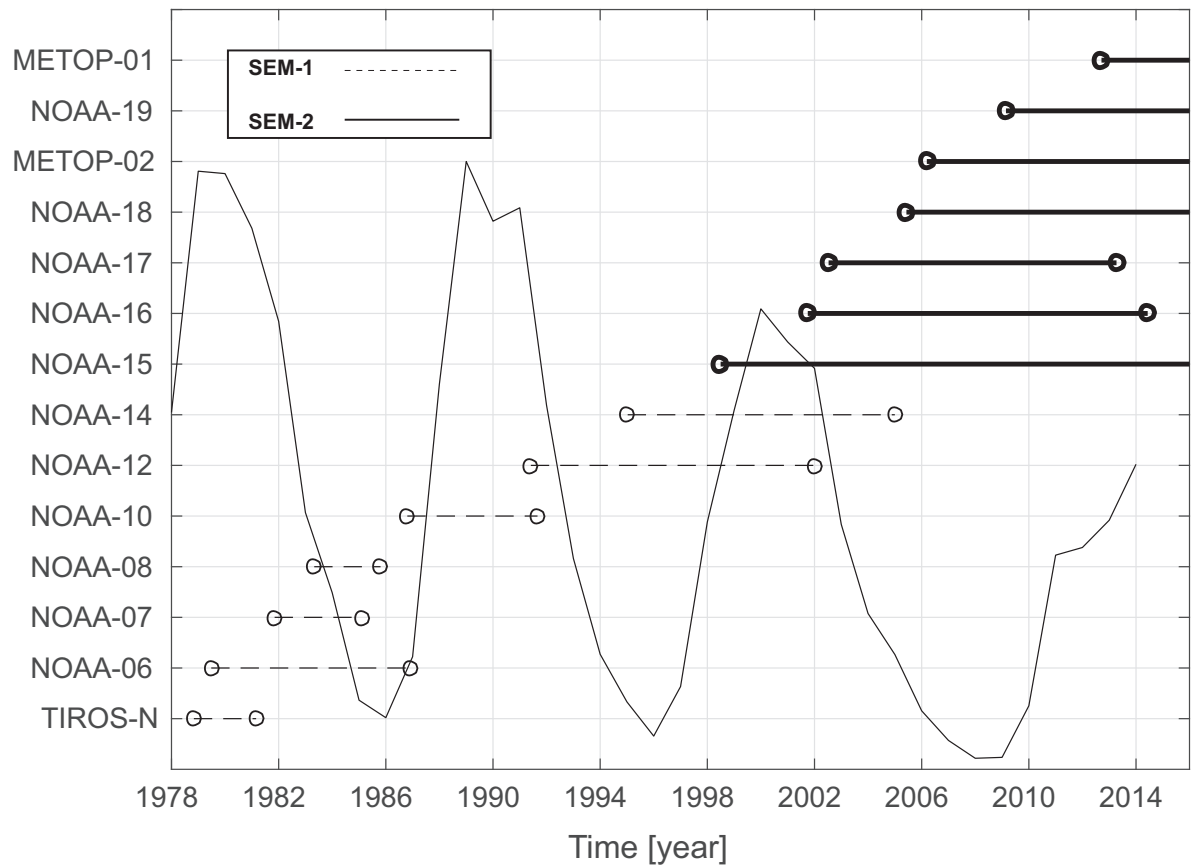

**Fig. 1.** The yearly sunspot number is shown as a solid black curve for reference. The operational lifetimes of the MEPED instrument on the SEM-1 satellites are plotted as dashed lines, and the SEM-2 satellites in thick solid lines.

Due to the general anisotropy in the proton flux the 90° detector is exposed to more radiation than the 0° detector throughout most of an orbit. The mean daily flux of the 90° detector is therefore initially higher than that of the 0° detector. However, the higher radiation dose will also degrade the 90° detector faster than the 0° detector, and, as seen in Figure 2b, the flux ratio between the two detectors becomes more and more unified until the 0° detector achieves a higher flux than the 90° detector in 2007.

The first extensive work of calibrating the entire MEPED proton dataset was done by Asikainen & Mursula (2011). An improved calibration was published the year after (Asikainen et al. 2012). Sandanger et al. (2015) used a different approach, utilizing a larger part of the available dataset, and accounting for alternating morning and evening orbits of the SEM-2 satellites (Magnetic Local Time effects).

Common for these studies is the technique used to quantify the degradation of a detector. When possible by criteria set by the authors, an integral energy spectrum from a newly launched satellite is compared to the integral energy spectrum from the degraded satellite. The detectors on the two satellites are assumed to measure the same particle population, and the new detector is expected to measure the real, non-degraded spectrum of particles. From this the increased energy thresholds of the degraded detector can be determined. The ratio of new energy threshold to the nominal energy threshold of a channel is called the $\alpha$ factor, or correction factor, for the energy channel in question. Details of comparing energy spectra and obtaining correction factors are described both in Asikainen & Mursula (2011) and Sandanger et al. (2015). The criteria for when a comparison between satellites can be made, which are the main distinctions between the methods, are summarized in Table 2.

The different criteria set by Asikainen & Mursula (2011) and Sandanger et al. (2015) provide different results. For example, NOAA-15 and NOAA-19 were excluded from calibration by the criteria set on separation in space by Sandanger et al. (2015), whereas Asikainen & Mursula (2011) were not able to calibrate the higher energy channels P4 and P5. Due to the small uncertainties in the $\alpha$ factors, and the consistent way of treating the time evolution of the factors, the results by Sandanger et al. (2015) are chosen for this study. The $\alpha$ factors calculated by comparing satellites from Sandanger et al. (2015) are presented here in Table 3. For each possible comparison, an $\alpha$ was calculated for each of the energy channels P1–P5, denoted as $\alpha_1$–$\alpha_5$ in the table.

Sandanger et al. (2015) proposed an iterative method to calculate the time evolution of the $\alpha$-factors from Table 3 with a monthly resolution. The $\alpha$ factors are calculated as a function of the accumulated *corrected* integral flux above a chosen energy threshold. We will make use of this method to get a reasonable estimation of the accumulated corrected fluxes, and show that there is a close relationship between the accumulated corrected fluxes and the $\alpha$ factors, as was also noted by Asikainen et al. (2012).

## 5. Predicting future degradation

The detectors will be degraded by the particles they measure. This implies that the degradation of the different detectors and the $\alpha$ factors will be dependent on the number of protons hitting the detectors. Assuming that the detectors are identical, the degradation process should influence all the detectors in the same way, independent of pointing direction and spacecraft. Therefore, if the correction factors found by Sandanger et al. (2015) are reliable, it should be possible to develop an analytical expression for the degradation valid for the MEPED proton detectors in both SEM-1 and SEM-2 satellites.

Sandanger et al. (2015) assume that the number of protons hitting the detector throughout its lifetime is responsible for raising the energy thresholds. Since the degradation raises the energy thresholds, it will not be long before the detectors are unable to measure, for example, 30 keV protons in the P1 channel. Sandanger et al. (2015) chose to use a range of energies to represent the particle population responsible for the degradation which even the most degraded of the satellites they consider is able to measure; [60, 167, 360, 1050, 2300] keV. Each monthly integral proton spectrum is corrected by the appropriate α factor, and the fluxes are accumulated at the given thresholds. For a detailed description of the procedure we refer the reader to Sandanger et al. (2015) Section 6 and Figure 6.

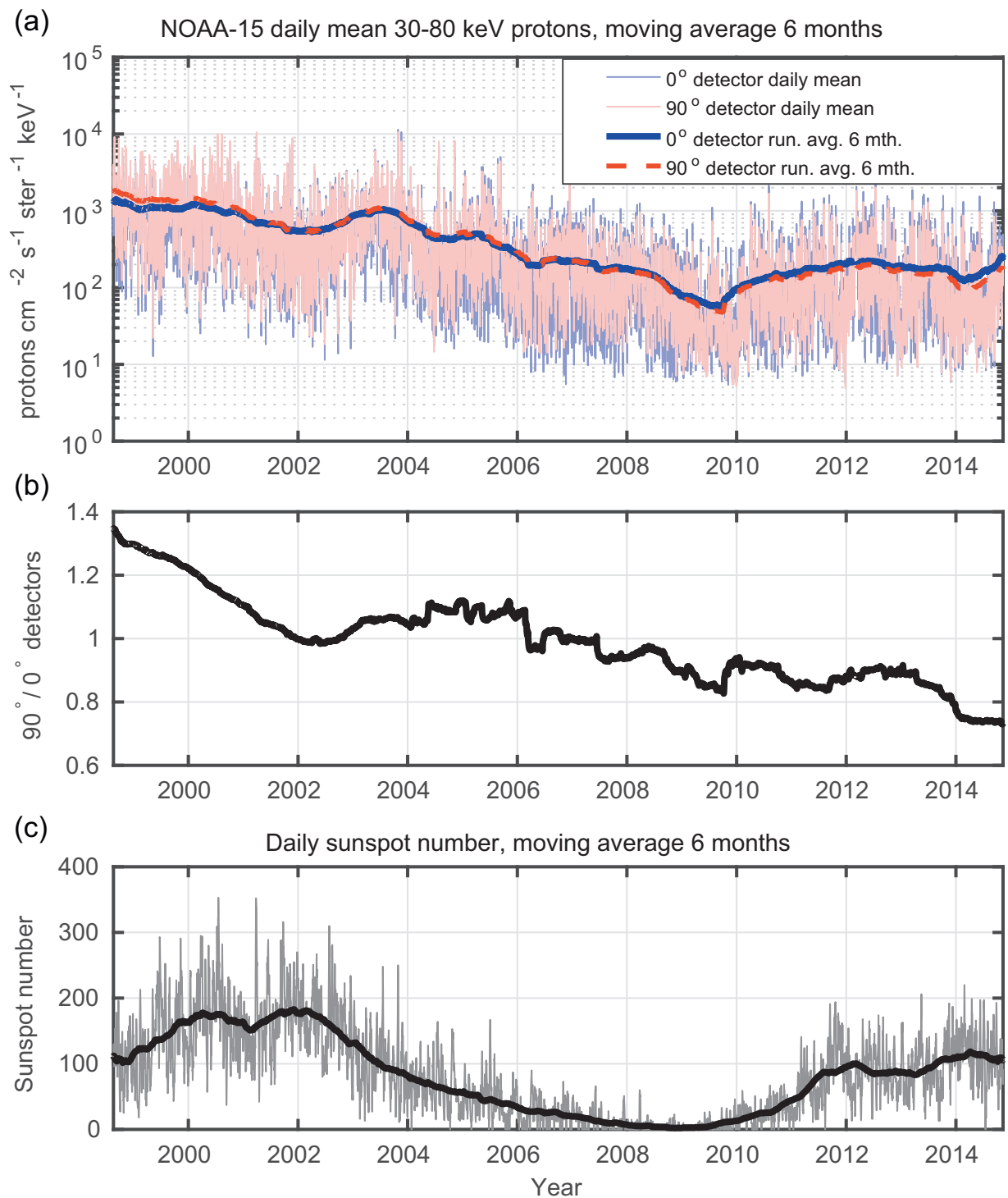

**Fig. 2.** (a) Daily mean uncorrected proton differential flux as measured by NOAA-15 P1 channel (30–80 keV) shaded in black. A running mean with a 6-month window is applied and plotted in solid and dashed lines on top. (b) 6-month running mean of 90° detector flux divided by that of the 0° detector. (c) The daily sunspot number shaded in grey. A moving average with a window of 6 months is applied to the data, and plotted in solid line on top. All panels show the period August 1998 to December 2014, each tick on the *x*-axis shows the 1st of January that year.

**Table 2.** Comparison of calibration methods and criteria.

| | Asikainen & Mursula (2011) and Asikainen et al. (2012) | Sandanger et al. (2015) |
|---|---|---|
| α Factors provided for channels[a] | P1–P3 | P1–P5 |
| Separation in space between satellites | <1° magnetic latitude and longitude at fofl[b] and <10% relative difference in *L*-value | Only compare satellites with <1 h separation in MLT |
| Separation in time between comparable satellites | <30 s | – |
| Data used | Instantaneous measurements (2 s resolution) | Mean value of one month of data from whole orbit |
| Number of months data from new satellite is considered non-degraded | 5 | 12 |
| Method for obtaining time evolution | Linear, 2nd order, 3rd order, PCHIP[c] fit with α vs. time, or as a function of *ap* index | Iteration method using α vs. accumulated corrected integral flux |

[a] P6 is excluded in all studies due to relativistic electrons contaminating the measurements.
[b] foot of field line (fofl) confining each satellite.
[c] Piecewise Cubic Hermite Interpolation Polynomial (PCHIP), shape preserving interpolation.

**Table 3.** The calibration α factors with standard deviations calculated by Sandanger et al. (2015).

| Satellite | Year | Mean α (standard deviation) | | | | |
|---|---|---|---|---|---|---|
| | | $\alpha_1$ | $\alpha_2$ | $\alpha_3$ | $\alpha_4$ | $\alpha_5$ |
| | | 0° detector | | | | |
| NOAA-16 | 2005 (Feb) | 1.57 (0.08) | 1.65 (0.08) | 1.22 (0.05) | 1.12 (0.06) | 1.08 (0.08) |
| NOAA-17 | 2007 (Jul) | 1.37 (0.06) | 1.59 (0.07) | 1.19 (0.02) | 1.07 (0.01) | 1.03 (0.01) |
| NOAA-17 | 2013 (Apr) | 1.53 (0.05) | 1.82 (0.10) | 1.32 (0.07) | 1.27 (0.09) | 1.16 (0.08) |
| NOAA-18 | 2009 (Sep) | 1.06 (0.04) | 1.14 (0.05) | 1.07 (0.01) | 1.16 (0.01) | 1.17 (0.01) |
| MetOp-02 | 2013 (Apr) | 1.13 (0.07) | 1.27 (0.10) | 1.09 (0.07) | 1.04 (0.06) | 1.05 (0.06) |
| | | 90° detector | | | | |
| NOAA-16 | 2005 (Feb) | 1.89 (0.10) | 1.86 (0.10) | 1.31 (0.02) | 1.08 (0.03) | 1.03 (0.03) |
| NOAA-17 | 2007 (Jul) | 1.65 (0.08) | 1.94 (0.07) | 1.46 (0.02) | 1.25 (0.01) | 1.21 (0.01) |
| NOAA-17 | 2013 (Apr) | 1.94 (0.06) | 2.10 (0.10) | 1.39 (0.09) | 1.22 (0.06) | 1.21 (0.05) |
| NOAA-18 | 2009 (Sep) | 1.11 (0.05) | 1.28 (0.06) | 1.12 (0.01) | 1.05 (0.01) | 1.03 (0.01) |
| MetOp-02 | 2013 (Apr) | 1.30 (0.09) | 1.47 (0.10) | 1.10 (0.06) | 0.92 (0.05) | 0.92 (0.05) |

We apply the method described by Sandanger et al. (2015). For every month in operation, we use all available data from the entire orbit and calculate a mean integral proton spectrum. Instead of using the energy thresholds as proposed by Sandanger et al. (2015) when accumulating the corrected fluxes, we performed a correlation analysis of their α factors versus the accumulated corrected fluxes for energies ranging from 90 keV to 2400 keV. The analysis resulted in maximum correlation for 90 keV independent of energy channel. A lower limit of 90 keV corresponds to an $\alpha = 3$ in the P1 channel. There is no physical limit for the maximum degradation at $\alpha = 3$, but we set the limit here as a tradeoff between allowing for a large degradation in the oldest detectors and keeping the energy threshold low to include the lower energy part of the spectrum in the accumulated flux. Sandanger et al. (2015) got a maximum $\alpha = 1.94$ for P1, while Asikainen et al. (2012) calculated a maximum $\alpha = 3.74$ for the NOAA-6 P1 channel. None of the other Asikainen et al. (2012) P1 factors were above 3 however, and we thus evaluate 90 keV as a suitable threshold. The resulting time evolutions of α as a function of accumulated corrected flux with $E > 90$ keV are plotted in Figure 3.

It might seem counterintuitive that also the α factors of the higher energy channels correlate best with the accumulated flux if the lower energy population is included. However, considering the high fluxes associated with low energy compared to high energy protons, it is likely that the main degradation in the silicon crystal occurs in the front end of the first detector. The low energy protons detected in P1–P3 ($E < 800$ keV) will deposit all their energy in the front end of the first silicon crystal. Going through this damaged region, the higher energy protons will also deposit a portion of their energy there. The change in the threshold for higher energy channels can therefore partly be ascribed to the damage caused by the lower energy protons. Together with considerable inter-correlation between the fluxes in the different energy channels, this might explain the strong correlation with the $E > 90$ keV proton fluxes for all the channels.

α factors from Table 3 are scattered versus the calculated accumulated corrected flux with $E > 90$ keV in Figure 4. The correlation coefficients for P1–P5 are $r = [0.84, 0.91, 0.86, 0.60, 0.55]$. For P4 and P5 we have removed the α factors of the MetOp-02 90° detector, which have $\alpha < 1$, when calculating the correlation. We do not believe that it is possible for the detectors to achieve lower energy thresholds than the nominal. The P1–P3 correlation coefficients are significant at the 5% level, while P4 at the 10% level. For P5, the $p$-value of the correlation coefficient is 0.12.

### 5.1. Linear regression

We want to express the degradation coefficient α as a linear function of the accumulated corrected flux with $E > 90$ keV (from here on abbreviated "acf"). We use the α factors from Table 3 and the calculated acf and do Ordinary Least Squares regression (OLS) on the data. We have removed all $\alpha < 1$ because of the non-physical implications of a negative degradation. The results of the OLS are summarized in Table 4.

The $p$-values for the estimated slopes and intercepts are calculated for a Student's $t$-test with $n - 2$ degrees of freedom, and $p$-value $<0.05$ indicates significant estimates for the parameters. The slopes are tested for significant difference from zero, while the intercepts are tested for significant difference from one. It would be impractical as well as non-physical to have $\alpha < 1$ for low accumulated fluxes as was found for the P1 channel, however, the intercepts were not found to be significantly different from one for any of the channels. Therefore, we instead perform a new regression where we set the intercept constant to $\alpha = 1$, which also reduces the degrees of freedom in the regression to $n - 1$. (In practice, we perform a Regression Through the Origin (RTO) after subtracting 1 from all α factors, see e.g. Eisenhauer 2003). The results from the RTO are given in Table 5.

We can compare the standard errors of the estimated slope in the two regressions, which is at least 50% lower in all channels for the RTO, indicating that this is the best of the two alternatives to fit a line to the data. Also, the $p$-values of the estimated slopes are smaller for the RTO, where all the estimated slopes were found to be significant to the 5% level. The regression lines found by RTO are shown in Figure 4 together with the scatter plot of α versus *acf*. A 95% prediction interval is displayed for the regression, and we use a constant intercept = 1.

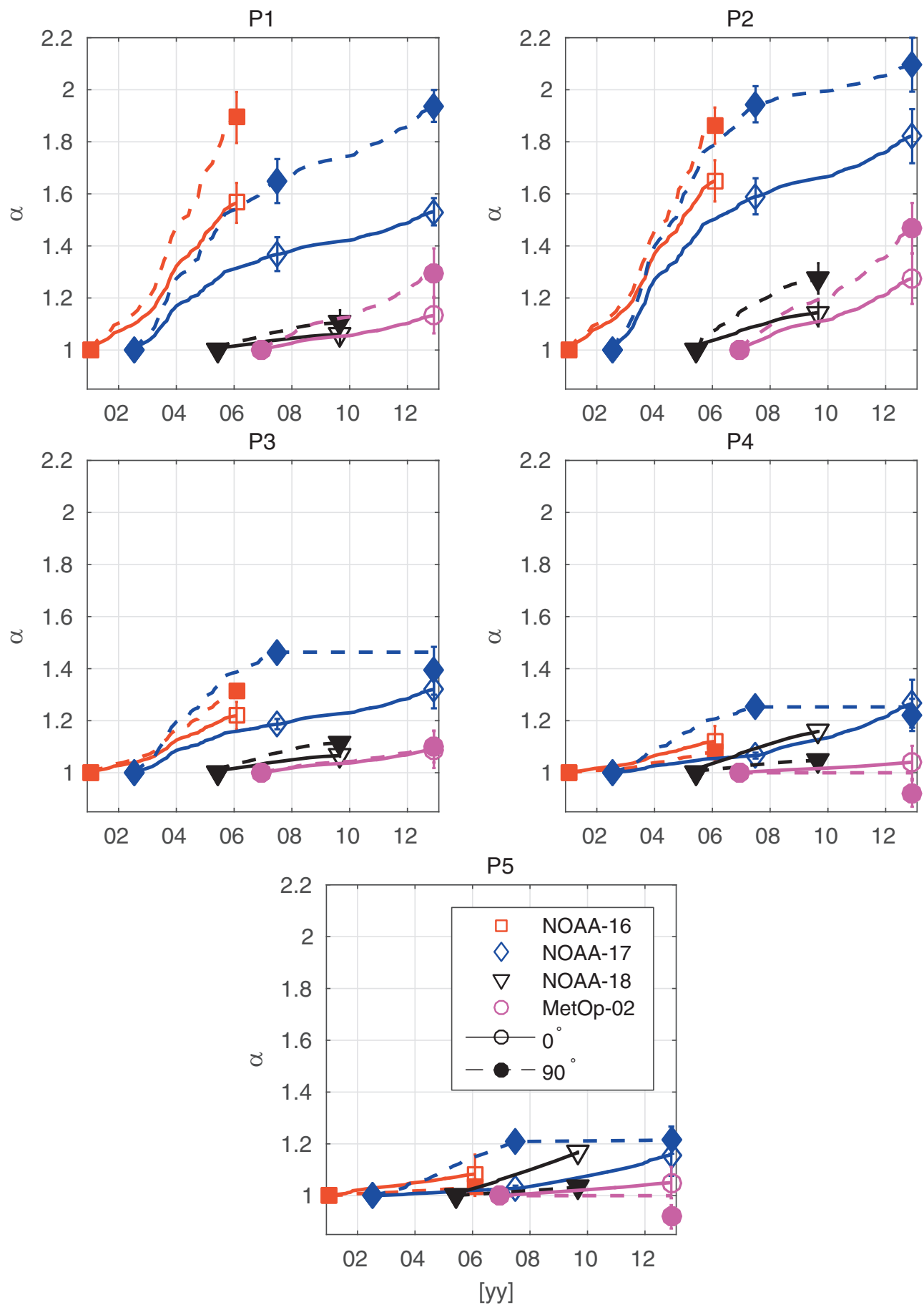

**Fig. 3.** α factors as a function of time for all satellites calibrated by Sandanger et al. (2015), both detectors. Upper row, left to right: P1 and P2 energy channels. Middle row, left to right: P3 and P4 energy channels. Bottom row: P5 energy channel. α is found as a function of the cumulative integral flux with $E > 90$ keV for all energy channels. The standard deviations of α are plotted as vertical lines.

From the RTO (Table 5), we get the following equations for P1–P5:

$$
\begin{aligned}
\alpha_{P1} &= 1.1 \times 10^{-6} acf + 1 \\
\alpha_{P2} &= 1.4 \times 10^{-6} acf + 1 \\
\alpha_{P3} &= 5.3 \times 10^{-7} acf + 1 \,, \\
\alpha_{P4} &= 3.0 \times 10^{-7} acf + 1 \\
\alpha_{P5} &= 2.3 \times 10^{-7} acf + 1
\end{aligned}
\tag{1}
$$

where *acf* is the accumulated corrected mean monthly integral flux with energy $E > 90$ keV.

The analytic expressions depend on the *acf* being a known parameter, which is not the case unless we already have a reliable estimate for the α factor of a detector. To achieve an estimated *acf*, we do a continuous month by month calculation as described in the next paragraph. This implies that we need to assume that the relationship found in Equation set (1) holds on time scales of a month, and not just on yearly basis.

The first step is to calculate the mean monthly integral flux from the uncorrected MEPED measurements for all months in operation. For the first month, we assume $\alpha = 1$ for all channels and make a continuous integral flux spectrum on the nominal energy thresholds with a Piecewise Cubic Hermite Interpolation Polynomial (PCHIP, a shape preserving interpolation routine) and numerically find the proton flux with $E > 90$ keV. The flux found is then applied in Equation set (1), and an $\alpha$ can be calculated for the next month. This calculated factor is used to correct the energy thresholds before constructing an integral flux spectrum for the next month, retrieving the flux with $E > 90$ keV and calculating the next $\alpha$ factor. The process is continued for as many months as there is data. For satellites previously corrected by Sandanger et al. (2015), we start by using the last provided factor in Table 3 and correcting the energy thresholds with these in the first step.

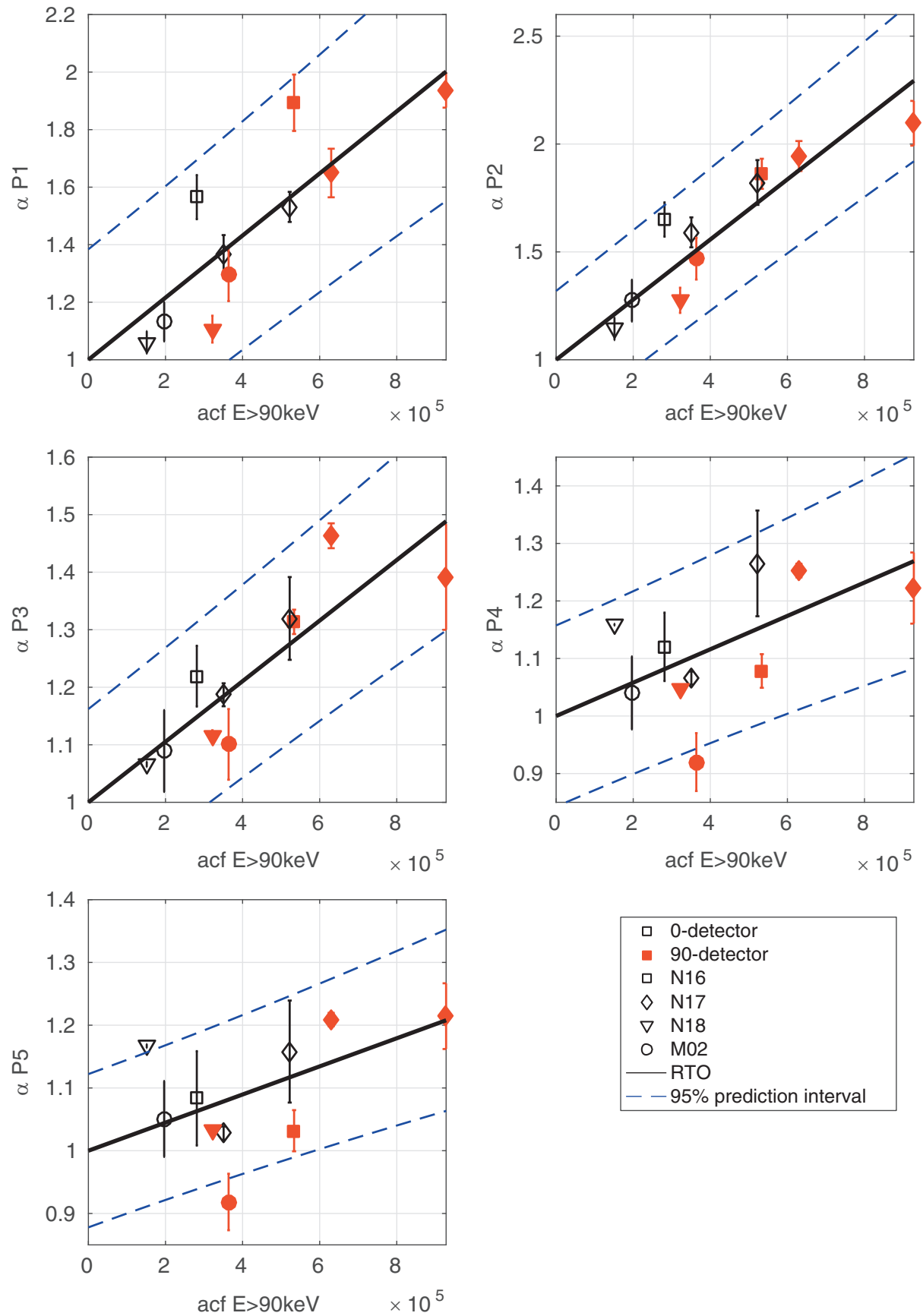

**Fig. 4.** α factors from Table 3 plotted versus the *acf* with energy $E > 90$ keV for all channels. The different satellites are plotted in different symbols, given in the legend. 0° and 90° detectors are distinguished by open black and solid red symbols, respectively. The NOAA-18 0° detector factors are removed from P4 and P5 in the regression. The best fit line found by regression is plotted in black, with accompanying 95% prediction intervals in dashed blue lines. The standard deviations of α are plotted as vertical lines.

**Table 4.** Ordinary least squares regression of α versus *acf*.

| Channel | Parameter | Estimate | Standard error | *p*-value |
|---------|-----------|----------|----------------|-----------|
| P1 | Intercept | 0.96 | 0.12 | 0.61 |
|    | Slope | $1.15 \times 10^{-6}$ | $2.6 \times 10^{-7}$ | $1.1 \times 10^{-3}$ |
| P2 | Intercept | 1.08 | 0.1 | 0.23 |
|    | Slope | $1.25 \times 10^{-6}$ | $2.1 \times 10^{-7}$ | $1.57 \times 10^{-4}$ |
| P3 | Intercept | 1.00 | 0.05 | 0.45 |
|    | Slope | $5.13 \times 10^{-7}$ | $1.1 \times 10^{-7}$ | $7.7 \times 10^{-4}$ |
| P4 | Intercept | 1.04 | 0.05 | 0.23 |
|    | Slope | $2.20 \times 10^{-7}$ | $1.1 \times 10^{-7}$ | 0.04 |
| P5 | Intercept | 1.03 | 0.05 | 0.28 |
|    | Slope | $1.78 \times 10^{-7}$ | $1.0 \times 10^{-7}$ | 0.06 |

**Table 5.** Regression with a constant intercept = 1, of $\alpha$ versus *acf*.

| Channel | Parameter | Estimate | Standard error | *p*-value |
|---------|-----------|----------|----------------|-----------|
| P1 | Slope | $1.1 \times 10^{-6}$ | $1.1 \times 10^{-7}$ | $2.4 \times 10^{-6}$ |
| P2 | Slope | $1.4 \times 10^{-6}$ | $9.3 \times 10^{-8}$ | $5.5 \times 10^{-8}$ |
| P3 | Slope | $5.3 \times 10^{-7}$ | $4.7 \times 10^{-8}$ | $7.1 \times 10^{-7}$ |
| P4 | Slope | $3.0 \times 10^{-7}$ | $5.0 \times 10^{-8}$ | $1.8 \times 10^{-4}$ |
| P5 | Slope | $2.3 \times 10^{-7}$ | $4.6 \times 10^{-8}$ | $4.5 \times 10^{-4}$ |

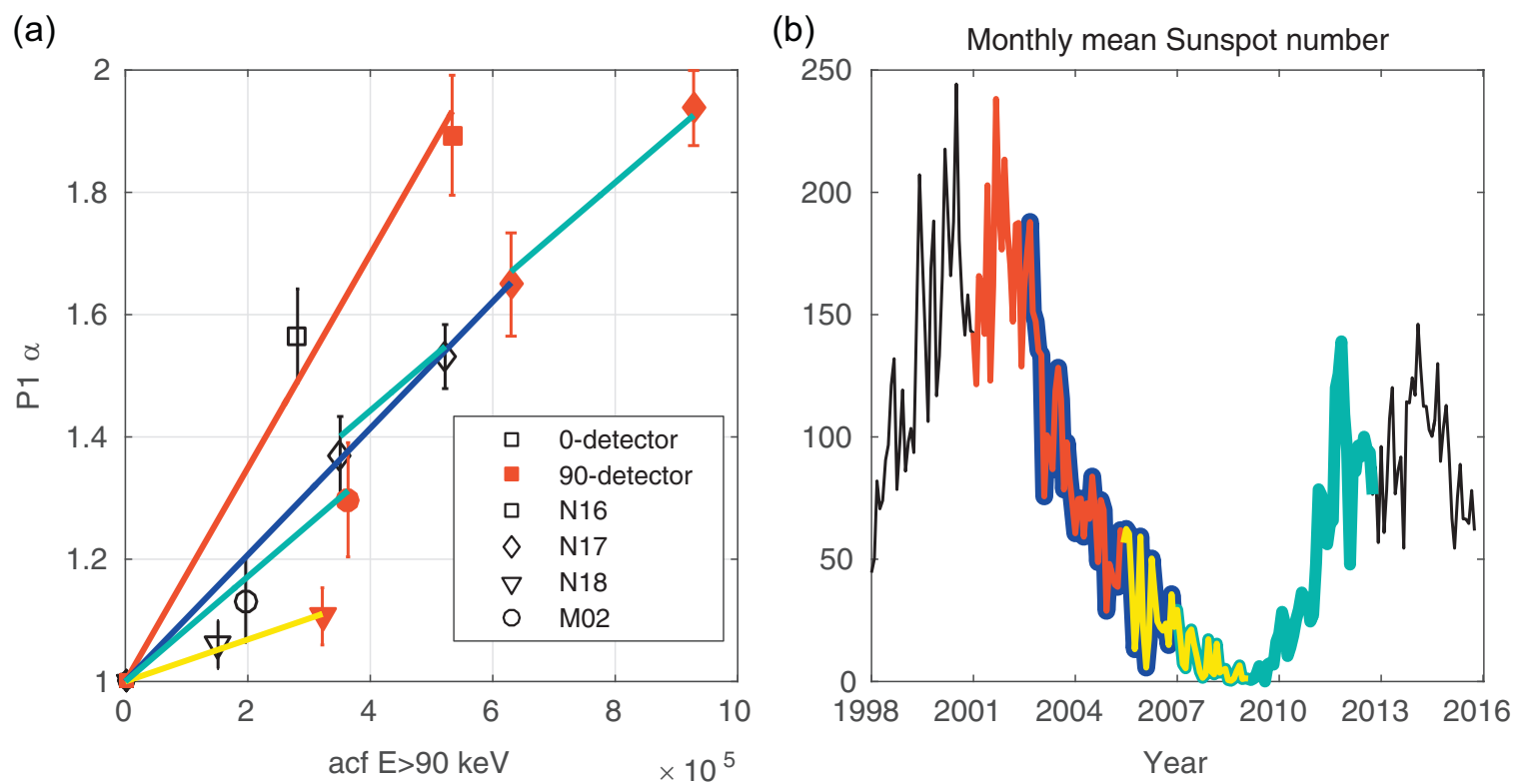

**Fig. 5.** (a) $\alpha$ factors for P1 plotted versus the *acf* ($E > 90$ keV). The standard deviations of $\alpha$ are plotted as vertical lines. (b) Monthly averaged sunspot number. Four calibration periods have been colour coded, and they are defined by the launch of new satellites. On the *x*-axis of the right panel each tick marks the 1st of January that year.

### 5.2. Introducing a dependence on *ap* index in the degradation

As a first approximation we have assumed that the degradation depends linearly on the accumulated corrected fluxes and proceeds similarly in all spacecraft. However, if we study the degradation of the satellites separately, we find that the degradation rate changes through the solar cycle, in particular for P1–P3. By degradation rate, we here mean $\Delta\alpha/\Delta acf$. That is, how much an $\alpha$ factor increases ($\Delta\alpha$) given a certain radiation dose ($\Delta acf$) over a given time, $\Delta t$. Galand & Evans (2000) noted that SEM-1 satellites launched near solar maximum seemed to degrade faster than those launched at the solar minimum, and that the solar history of the detectors was affecting the degradation rate of MEPED. We find the same effect in the SEM-2 MEPED instruments.

The effect is illustrated in Figure 5 for P1. Figure 5a shows the $\alpha$ factors versus *acf* for all the satellites calibrated by Sandanger et al. (2015), and the monthly sunspot number is plotted in Figure 5b. We have colour coded four periods which all correspond to the time between two $\alpha$ factors for one or more of the satellites as calculated by Sandanger et al. (2015). The slope between the *acf* and $\alpha$ for the different satellites changes through the solar cycle. The red period, which is closest to the solar maximum, has the steepest slope, while the yellow, which is closest to the solar minimum, has the gentlest slope. For example, approximately the same accumulated integral flux of :$3 \times 10^5$ [/cm² s ster] produces an $\alpha \approx 1.1$ in the NOAA-18 90° detector, and an $\alpha \approx 1.6$ in the NOAA-16 0° detector. $\Delta\alpha/\Delta acf$ is different through the solar cycle. We see the same feature in P2 and P3 as well (not shown). In addition, the two satellites MetOp-02 and NOAA-17, which have the turquoise period in common, have almost identical degradation rates during this period. This is an indication that there is some factor responsible for the different rates, and not just the detectors in the different satellites degrading in different manners. The definition of the red, blue, turquoise and yellow phases of the solar cycle is listed in Table 6.

**Table 6.** Definition of colour codes of solar cycle phases (see Fig. 5).

| Colour | From | To | $\alpha$ factors used |
|---|---|---|---|
| Red | Launch NOAA-16 (2001) | Launch NOAA-18 + 6 months (2005)[a] | $\alpha_1$ N16 |
| Green | Launch NOAA-17 (2002) | Launch MetOp-02 + 6 months (2007) | $\alpha_1$ N17 |
| Yellow | Launch NOAA-18 (2005) | Launch NOAA-19 + 6 months (2009) | $\alpha_1$ N18, |
| Turquoise | Launch MetOp-02 (2007) | Launch MetOp-01 + 6 months (2013) | $\alpha_1$ M02, $\alpha_2$ N17 |

[a] 6 months is added to the date of launch of the satellite defining the end of a period. This is because Sandanger et al. (2015) used one year of data from the new satellites to calibrate the old satellites and placed the calibration $\alpha$ factor in the middle of the respective 1-year period.

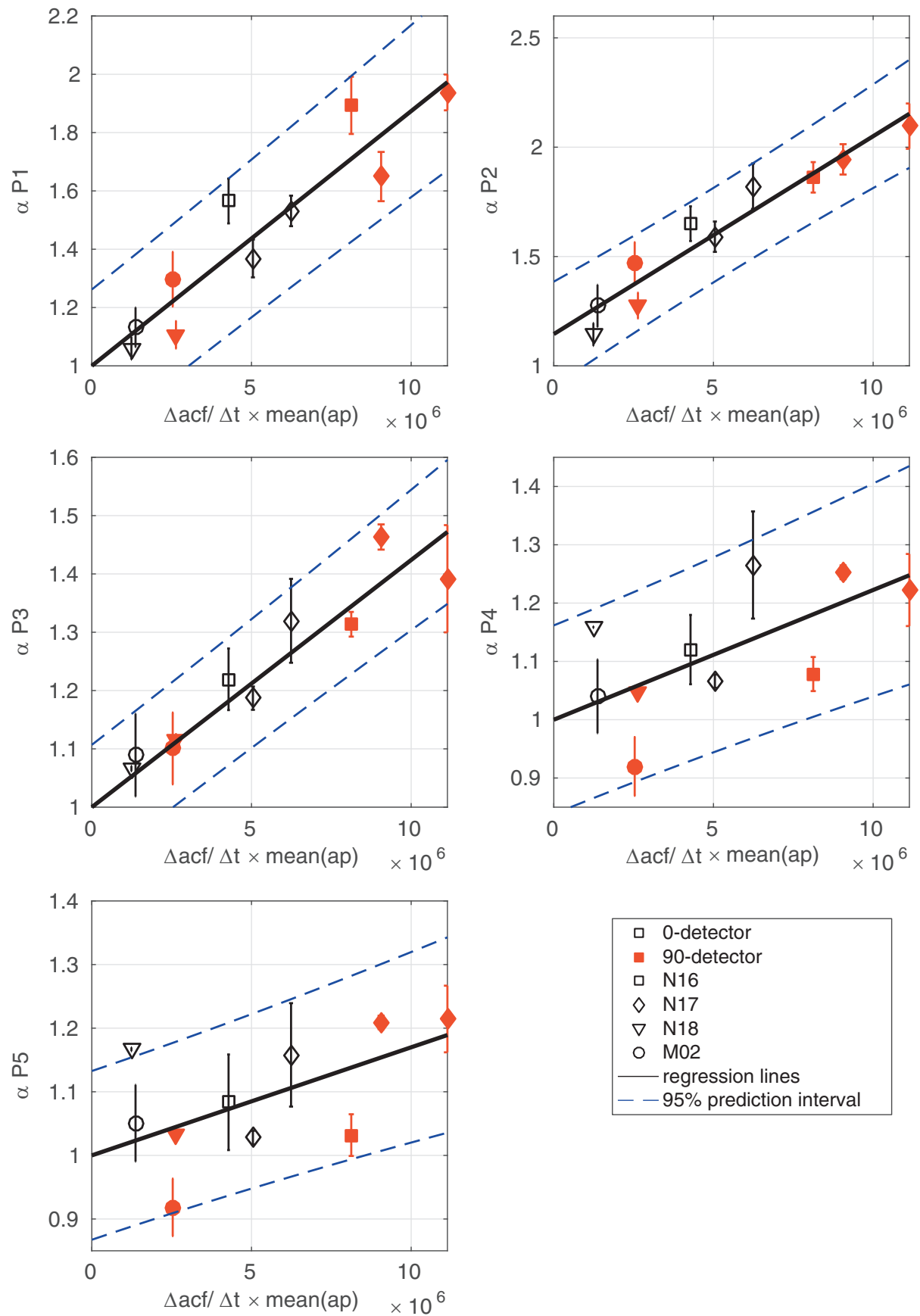

**Fig. 6.** $\alpha$ factors from Table 3 plotted versus the $\Delta acf/\Delta t$ ($E > 90$ keV) multiplied by the mean $ap$ index in each period. $\alpha < 1$ are removed when calculating the best fit in the regression. The equations for the best fit lines are presented in Eq. (2). A 95% prediction interval for the regressions is plotted in dashed blue lines. The standard deviations of $\alpha$ are plotted as vertical lines.

**Table 7.** Ordinary least squares regression of $\alpha$ versus $\Delta acf / \Delta t \times \overline{ap}$.

| Channel | Parameter | Estimate | Standard error | $p$-value |
|---------|-----------|----------|----------------|-----------|
| P1 | Intercept | 1.01 | 0.07 | 0.45 |
| | Slope | $8.6 \times 10^{-8}$ | $1.2 \times 10^{-8}$ | $4.7 \times 10^{-5}$ |
| P2 | Intercept | 1.15 | 0.05 | 0.01 |
| | Slope | $9.0 \times 10^{-8}$ | $8.7 \times 10^{-9}$ | $3.3 \times 10^{-6}$ |
| P3 | Intercept | 1.03 | 0.03 | 0.18 |
| | Slope | $3.9 \times 10^{-8}$ | $4.6 \times 10^{-9}$ | $1.6 \times 10^{-5}$ |
| P4 | Intercept | 1.06 | 0.05 | 0.15 |
| | Slope | $1.5 \times 10^{-8}$ | $7.9 \times 10^{-9}$ | 0.05 |
| P5 | Intercept | 1.05 | 0.05 | 0.17 |
| | Slope | $1.1 \times 10^{-8}$ | $7.5 \times 10^{-9}$ | 0.09 |

**Table 8.** Regression with a constant intercept = 1, of $\alpha$ versus $\Delta acf / \Delta t \times \overline{ap}$.

| Channel | Parameter | Estimate | Standard error | $p$-value |
|---------|-----------|----------|----------------|-----------|
| P1 | Slope | $8.7 \times 10^{-8}$ | $6.0 \times 10^{-9}$ | $7.4 \times 10^{-8}$ |
| P3 | Slope | $4.2 \times 10^{-8}$ | $2.5 \times 10^{-9}$ | $1.6 \times 10^{-8}$ |
| P4 | Slope | $2.3 \times 10^{-8}$ | $4.1 \times 10^{-9}$ | $2.9 \times 10^{-4}$ |
| P5 | Slope | $1.75 \times 10^{-8}$ | $3.8 \times 10^{-9}$ | $9.4 \times 10^{-4}$ |

One explanation of the observed variation in the degradation rate may be that it is caused by the proton population with energy $E < 90$ keV, which we have not included in our accumulated flux. To account for the variation, we weight the $\Delta acf / \Delta t$ by the mean $ap$ index over the respective period (where $\Delta t$ is the number of months in each of the coloured periods listed in Table 6). NOAA-17 is the only satellite calibrated over two periods. We calculate $\Delta acf / \Delta t$ separately for the two periods and multiply with the mean $ap$ (abbreviated $\overline{ap}$) for each period. The $\Delta acf_2 / \Delta t_2 \times \overline{ap}_2$, calculated for period 2, is added to the $\Delta acf_1 / \Delta t_1 \times \overline{ap}_1$ for period 1.

All $\alpha$ factors versus $\sum (\Delta acf / \Delta t \times \overline{ap})$ are plotted in Figure 6. The summation symbol indicates that we have to add together results from the two periods for NOAA-17. Inclusion of the $ap$ dependence improves the correlation coefficient in P1 from $r = 0.84$ to $r = 0.93$, in P2 from $r = 0.91$ to $r = 0.96$ and in P3 from $r = 0.86$ to $r = 0.95$. For P4 and P5 the correlation coefficients decrease slightly from $r = 0.60$ to $r = 0.58$ in P4, and from $r = 0.55$ to $r = 0.49$ in P5. P1–P3 and P5 correlations are significant to the 5% level, while P4 is significant to the 10% level. The fact that P4 and P5 are not much affected by the weighting with $ap$ index can be interpreted as support for the theory that the lower energy proton population is influencing the degradation of the P1–P3 channels. As mentioned before, the lower energy protons will deposit their energy in the front part of the detector, and thus be of lesser importance for the P4 and P5 channels, which measure particles with energies that penetrate deep into the detector.

We performed an OLS with $\alpha$ versus $\sum (\Delta acf / \Delta t \times \overline{ap})$. The estimated parameters for P1–P5 are summarized in Table 7. The regression lines for P1, P3, P4 and P5 are not found to have intercepts significantly different from $\alpha = 1$. We therefore proceeded to a RTO for these channels, the results are presented in Table 8. Since the intercept is significant for P2 in the OLS, we select this as the best fit. For the four other channels, we use the fit found by RTO. The slopes of P1, P3, P4 and P5 have approximately 50% smaller standard error in the RTO compared to the OLS. The best fit lines are summarized in Equation set (2) and plotted in Figure 6 together with a 95% prediction interval and the $\alpha$ factors by Sandanger et al. (2015). We can now express a refined degradation dependency:

$$\alpha_{P1} = 8.7 \times 10^{-8} \sum (\Delta acf / \Delta t \times \overline{ap}) + 1$$
$$\alpha_{P2} = 9.0 \times 10^{-8} \sum (\Delta acf / \Delta t \times \overline{ap}) + 1.15$$
$$\alpha_{P3} = 4.2 \times 10^{-8} \sum (\Delta acf / \Delta t \times \overline{ap}) + 1 \quad , \quad (2)$$
$$\alpha_{P4} = 2.2 \times 10^{-8} \sum (\Delta acf / \Delta t \times \overline{ap}) + 1$$
$$\alpha_{P5} = 1.7 \times 10^{-8} \sum (\Delta acf / \Delta t \times \overline{ap}) + 1$$

where the summation should be done over the number of periods chosen. That is, the $\alpha$ factor at the end of period 1 is $\alpha_1 : \Delta acf_1 / \Delta t_1 \times \overline{ap}_1$, that after period 2 is $\alpha_2 : \Delta acf_1 / \Delta t_1 \times \overline{ap}_1 + \Delta acf_2 / \Delta t_2 \times \overline{ap}_2$, and so on.

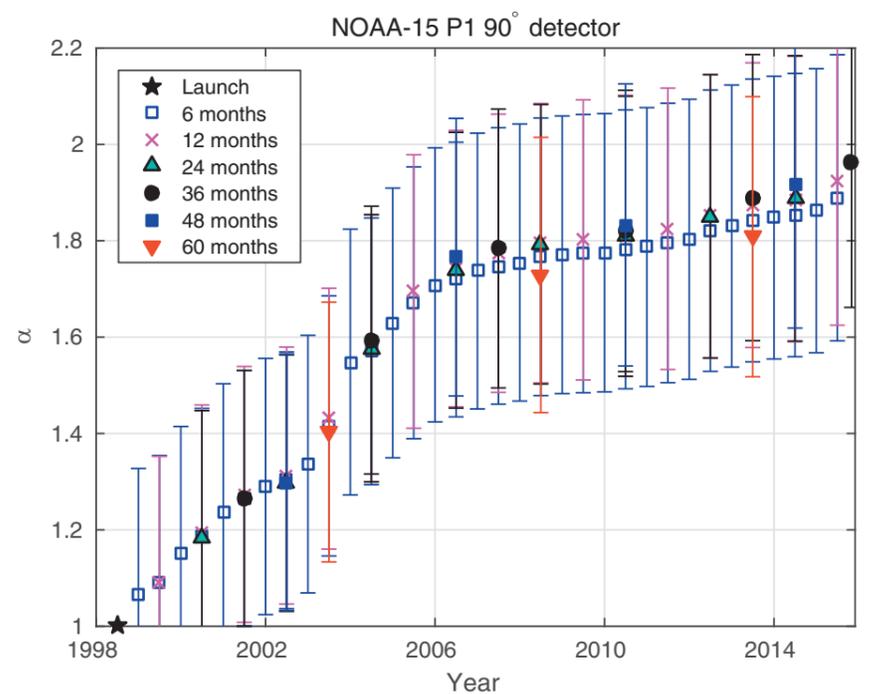

**Fig. 7.** $\alpha$ for the 90° detector of NOAA-15 calculated using different number of months between each factor. The 95% prediction interval is used to calculate errors.

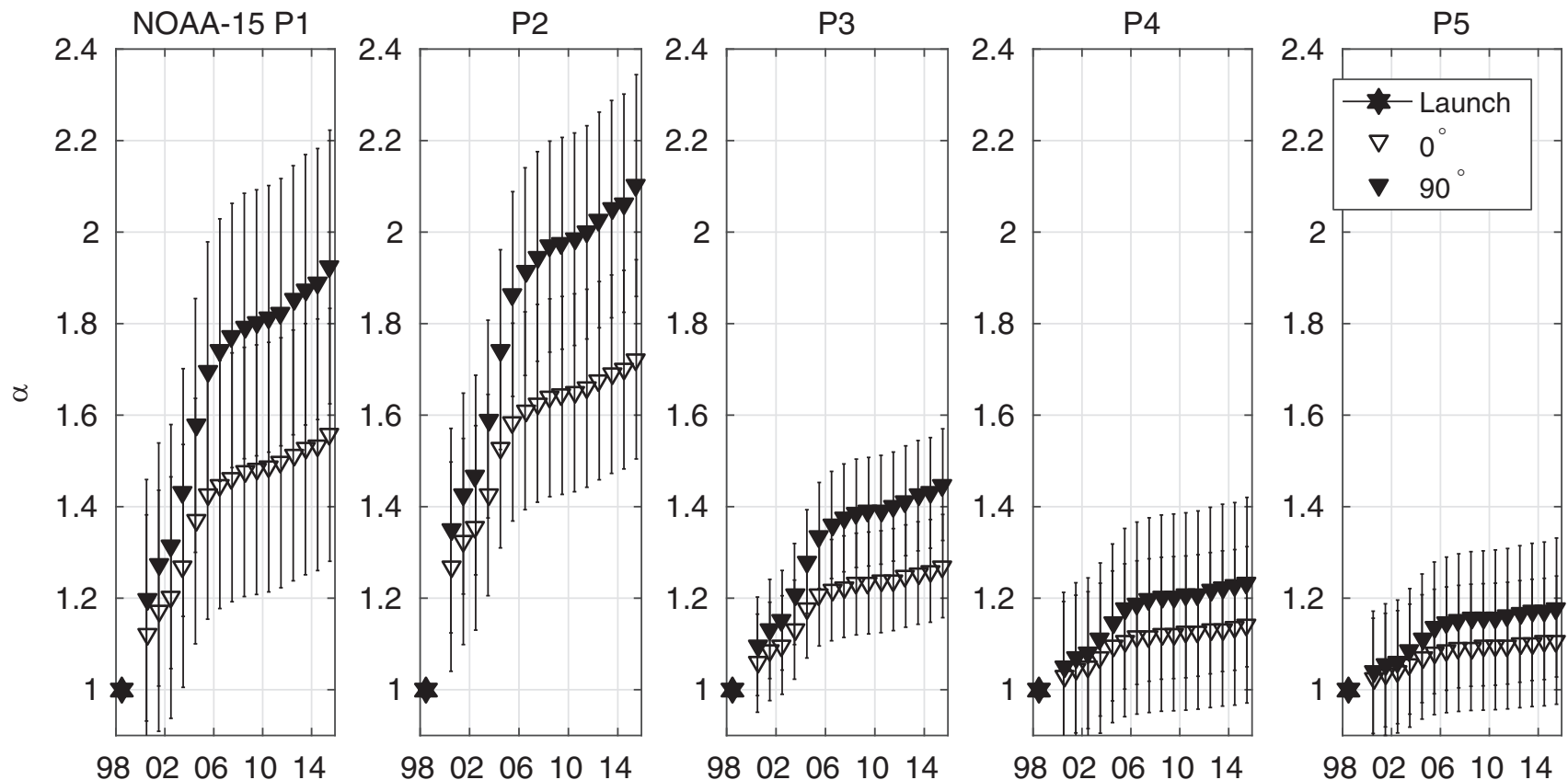

**Fig. 8.** α factors for NOAA-15 with 12 months in-between, calculated as described in Section 5.3. The 0° detector is shown as open triangles, and the 90° as filled triangles. The estimated error is calculated based on the 95% prediction interval.

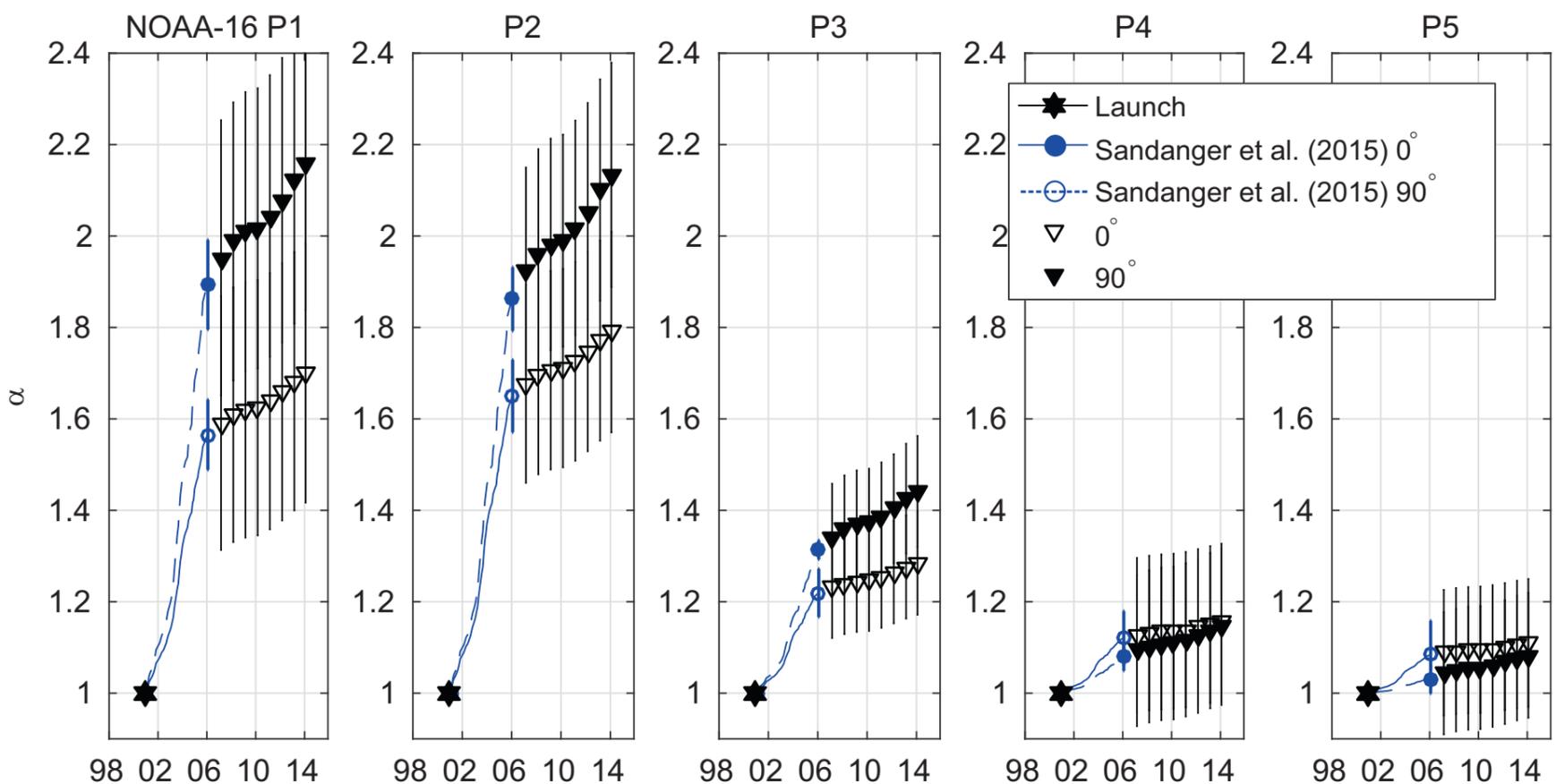

**Fig. 9.** α factors for NOAA-16 with 12 months in-between, calculated as described in Section 5.3. The 0° detector is shown as open triangles, and the 90° as filled triangles. The estimated error is calculated based on the 95% prediction interval. Factors from Sandanger et al. (2015) are shown as open (0°) and filled (90°) blue circles. Monthly α factors calculated with their method for E > 90 keV are displayed as blue solid (0°) and dashed (90°) lines.

Equation set (1) calculated α factors with a non-changing slope between α and *acf*. Equation set (2) weights the change in *acf*, that is Δ*acf*, through a time period Δ*t* with the mean *ap* index over the same time period. In this way we introduce an element which will increase the radiation dose efficiency in degrading the detectors (producing a Δα) in periods where the geomagnetic activity is high.

### 5.3. Calibration of all satellites

We now have the tools needed to calculate calibration factors for all the degraded detectors. The first step is to calculate monthly mean integral fluxes from the degraded data for the satellite we want to calibrate. We assume that α = 1 for all energy channels in the first month of operation, and thus the corrected flux is equal to the uncorrected flux the first month. For the satellites already calibrated by Sandanger et al. (2015), we start the use of our method in the first month after their last calibration factor and use their α to correct the energy spectrum. We use Equation set (1) to calculate a first approximation of monthly α and correct the energy thresholds by these preliminary α factors. The integral flux of $E > 90$ keV is found numerically in each month and accumulated in the process, that is, we calculate the *acf*.

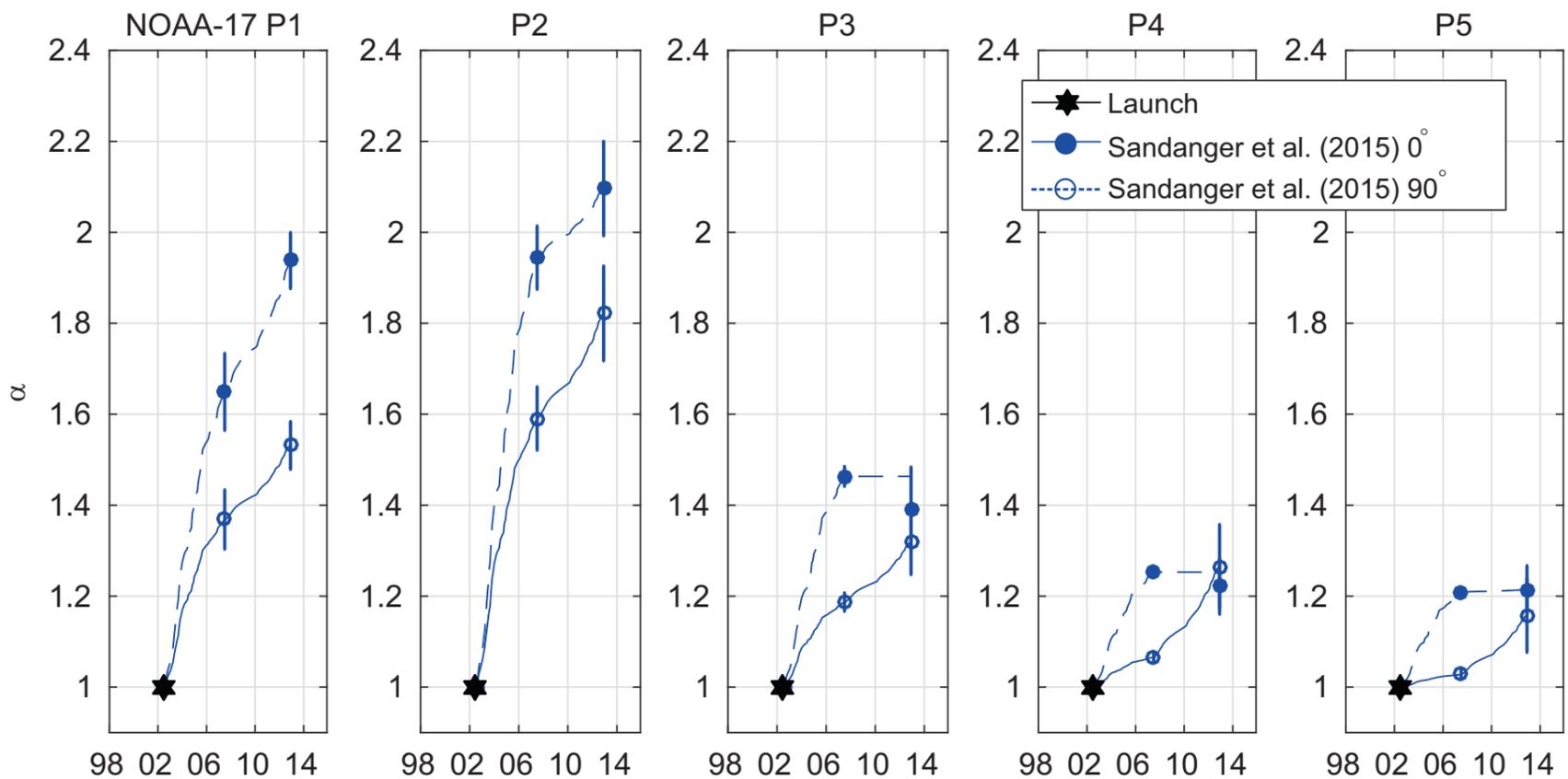

**Fig. 10.** NOAA-17 α factors from Sandanger et al. (2015) are shown as open (0°) and filled (90°) blue circles. Monthly α factors calculated with their method for $E > 90$ keV are displayed as blue solid (0°) and dashed (90°) lines.

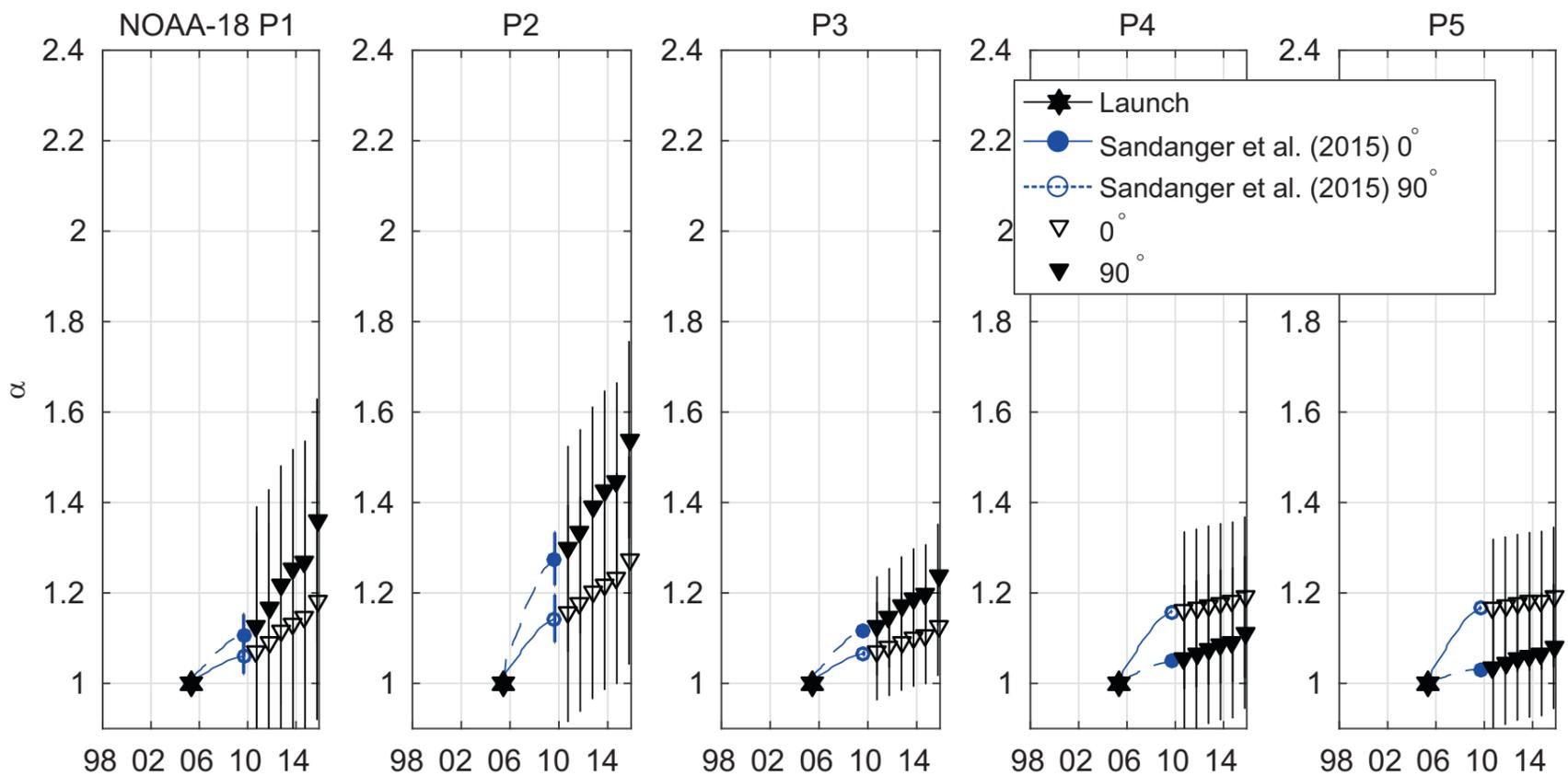

**Fig. 11.** α factors for NOAA-18 with 12 months in-between, calculated as described in Section 5.3. The 0° detector is shown as open triangles, and the 90° as filled triangles. The estimated error is calculated based on the 95% prediction interval. Factors from Sandanger et al. (2015) are shown as open (0°) and filled (90°) blue circles. Monthly α factors calculated with their method for $E > 90$ keV are displayed as blue solid (0°) and dashed (90°) lines.

After we have achieved an estimate for the *acf* for all the months in operation, we choose an interval of appropriately many months (e.g. $\Delta t = 12$) and find the mean *ap* index in each of the intervals of length $\Delta t$. We multiply the $\Delta acf/\Delta t$ by the mean *ap* index in each interval and add the results together. Lastly, we use the sum of $\Delta acf/\Delta t \times \overline{ap}$ to calculate α by Eq. (2). We use the new estimation of α factors from Eq. (2) to go back and correct the uncorrected monthly energy spectra and calculate the *acf* again. We can use the new *acf* to calculate $\Delta acf / \Delta t \times \overline{ap}$ for the $\Delta t = 12$-month intervals, and from Equation set (2) estimate α again. The iteration process is repeated until the α factors calculated from Eq. (2) converge, which happens after 3–5 iterations.

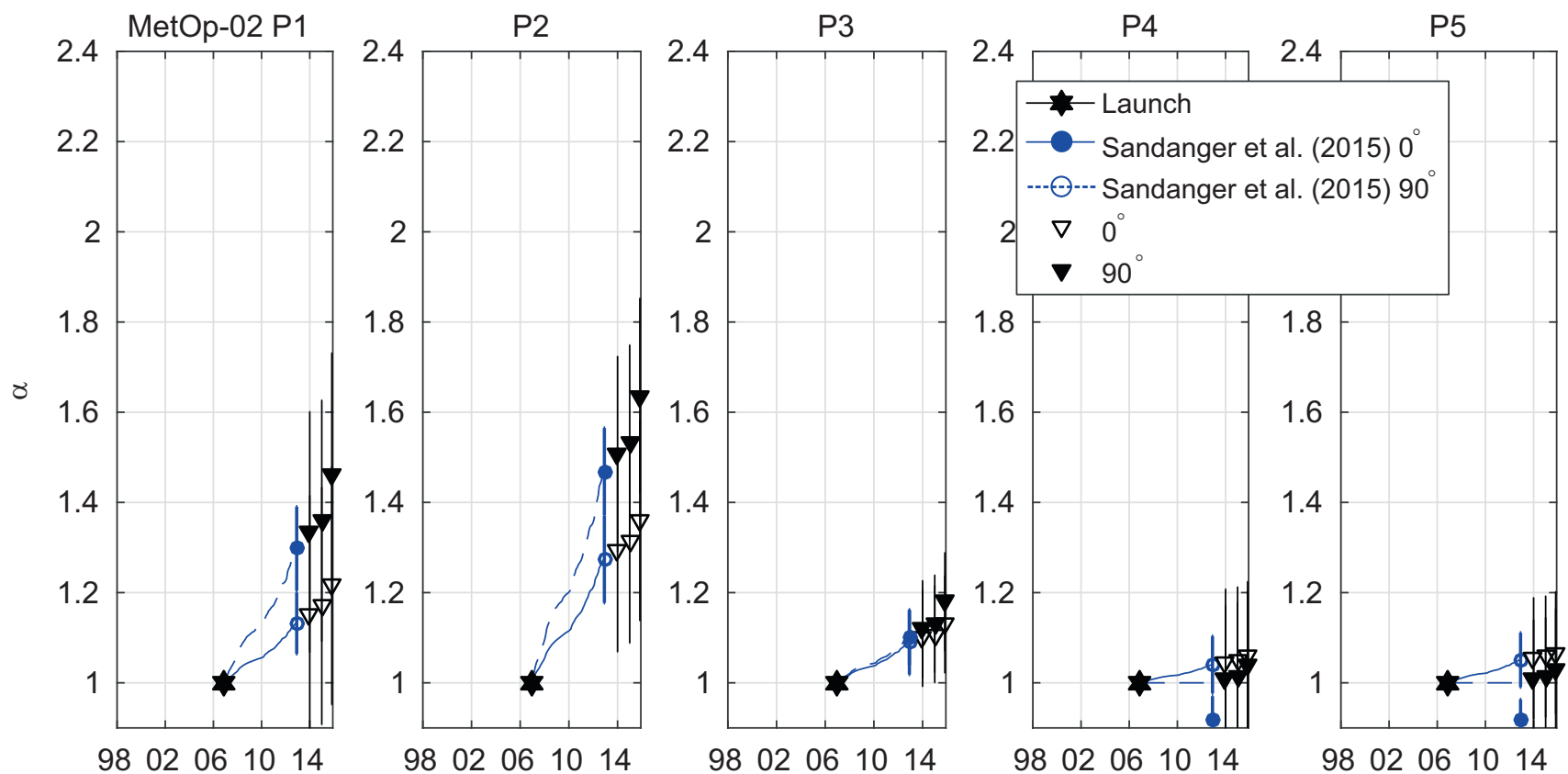

**Fig. 12.** α factors for MetOp-02 with 12 months in-between, calculated as described in Section 5.3. The 0° detector is shown as open triangles, and the 90° as filled triangles. The estimated error is calculated based on the 95% prediction interval. Factors from Sandanger et al. (2015) are shown as open (0°) and filled (90°) blue circles. Monthly α factors calculated with their method for $E > 90$ keV are displayed as blue solid (0°) and dashed (90°) lines.

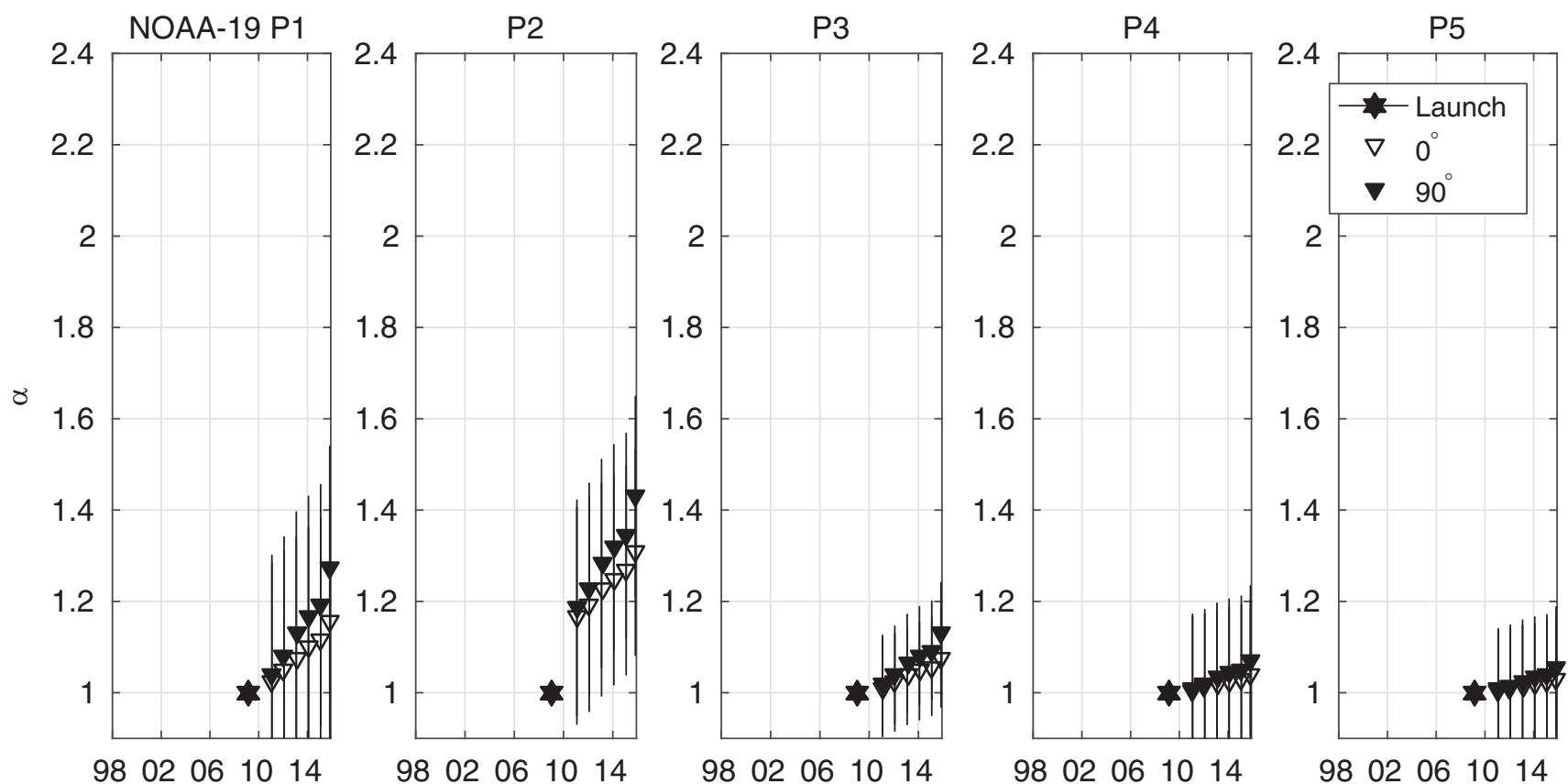

**Fig. 13.** α factors for NOAA-19 with 12 months in-between, calculated as described in Section 5.3. The 0° detector is shown as open triangles, and the 90° as filled triangles. The estimated error is calculated based on the 95% prediction interval.

It is not crucial to use a period of 12 months, as is illustrated in Figure 7 for NOAA-15's 90° detector. However, we encourage the use of intervals on the scale of years, rather than months. It seems that 6 months is a lower limit for the model to give consistent results. This is probably because the $ap$ index varies in a more random fashion on the scale of one to a few months than on yearly basis. It is the trend of the $ap$ index through the solar cycle rather than the monthly fluctuations that works as a weighting factor. In practice, by our method one can retrieve an estimated correction factor for a satellite in any chosen year directly (as shown in Fig. 7).

α factors for all satellites are calculated and plotted in Figures 8–14. The NOAA-17 MEPED lifetime ended in 2013, and we cannot extend the time series of α factors further than Sandanger et al. (2015). For NOAA-16, NOAA-18 and MetOp-02 we have plotted the α factors by Sandanger et al. (2015) on a monthly basis in blue (although, calculated as a function of *acf* with $E > 90$ keV for all channels, as opposed to their varying thresholds given here in Sect. 5). These are calculated using their method where each satellite and detector have an individually best fit slope between α and the corrected accumulated flux. In Figure 5 we showed how the degradation rate ($\Delta\alpha/\Delta acf$) varied for the satellites calibrated by Sandanger et al. (2015). The α factors presented in blue in Figures 9–12 thus have this variation included in them. α factors presented as black triangles are calculated using the method presented in this paper (filled triangles represent the 90° detector, open triangles the 0° detector). The error estimates were calculated from the 95% prediction intervals shown in Figure 6. The estimated errors are as large as ±0.27 in P1. The α factors estimated by satellite comparison in Sandanger et al. (2015) had smaller errors and are the preferred method to use when possible. We have here presented a model based on their results, and thus we expect the errors in our model estimated α factors to be larger. However, the correlations of 0.93, 0.96 and 0.95 in P1, P2 and P3 are encouraging in that the data our model is built on behaves consistently. The respective $R^2$ for the P1–P3 regressions (Eq. (2)) are 0.91, 0.93 and 0.93. For both P4 and P5, the regression $R^2$ is 0.85.

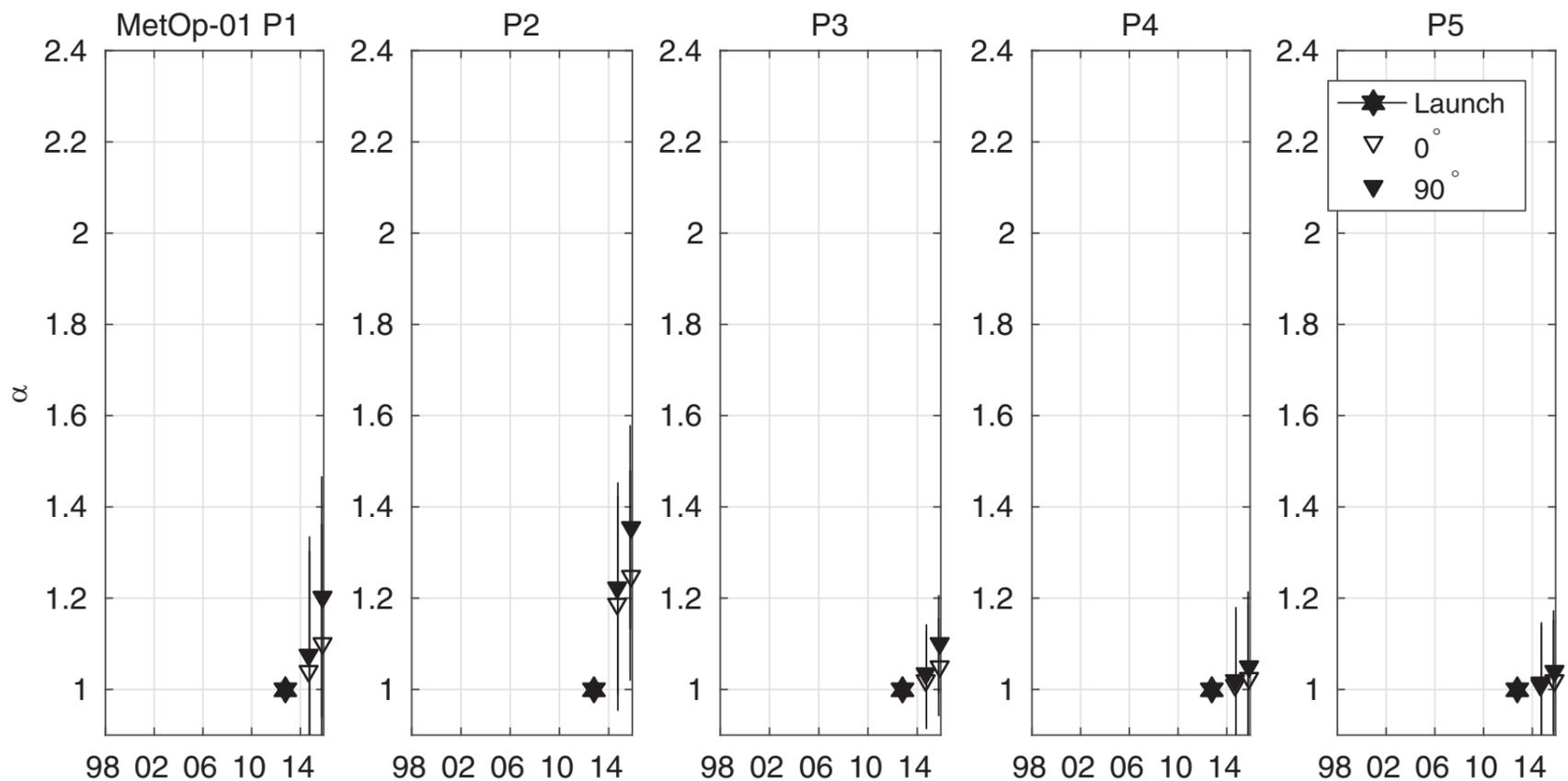

**Fig. 14.** α factors for MetOp-01 with 12 months in-between, calculated as described in Section 5.3. The 0° detector is shown as open triangles, and the 90° as filled triangles. The estimated error is calculated based on the 95% prediction interval.

# 6. Summary

There is a significant relationship between the correction α factors of all the MEPED proton channels P1–P5 and the accumulated corrected proton fluxes with $E > 90$ keV independent of the pointing direction and spacecraft. However, there is a variation in the degradation rate ($\Delta\alpha/\Delta acf$) through the solar cycle which may be caused by the variation of the flux of lower energy protons not included in the accumulated corrected flux (*acf*). For P1–P3 the correlation improves when the accumulated flux is weighted by the mean *ap* index. The *ap* index measures geomagnetic activity and can be used as a measure of how the particle population in the magnetosphere varies over the course of a solar cycle. We exploit this to introduce a weighting factor to go from the non-varying relationship between α and *acf* presented in Equation set (1), to the refined Equation set (2). Our model is based on correction factors from the period 2005 to 2013. There may be larger uncertainties connected with the model when applying it in the inclining and maximum phase of the previous solar cycle, which was stronger than the current cycle. Also, our model is based on good statistics for the evolution of α for satellites up to approximately 6 years of age. For higher age and degradation, the model is based on only NOAA-17. When MetOp-C is launched additional correction factors for MetOp-02 and MetOp-01 can be calculated using the Sandanger et al. (2015) method. This will improve the model. Finally, other methods for calibrating NOAA-15 should be investigated. This would be helpful in describing degradation for a stronger solar maximum than observed in the current solar cycle.

*Acknowledgements.* The authors thank the reviewers for helpful suggestions and comments. This study was supported by the Research Council of Norway/CoE under Contract 223252/F50. The authors thank D. S. Evans and the NOAA's National

Geophysical Data Center (NGDS) for providing NOAA data. We acknowledge the use of NASA/GSFC's Space Physics Data Facility's OMNIWeb service and OMNI data. We thank the SIDC team, World Data Center for the Sunspot Index, Royal Observatory of Belgium, Monthly Report on the International Sunspot Number, online catalogue of the sunspot index: http://www.sidc.be/sunspot-data/. The editor thanks Timo Asikainen and an anonymous referee for their assistance in evaluating this paper.

## References

Asikainen, T., and K. Mursula. Recalibration of the long-term NOAA/MEPED energetic proton measurements. *J. Atmos. Sol. Terr.Phys.*, **73**, 335–347, 2011, DOI: 10.1016/j.jastp.2009.12.011.

Asikainen, T., K. Mursula, and V. Maliniemi. Correction of detector noise and recalibration of NOAA/MEPED energetic proton fluxes. *J. Geophys. Res.*, **117 (A9)**, A09204, 2012, DOI: 10.1029/2012JA017593.

Davis, G. History of the NOAA satellite program. *J. Appl. Remote Sens.*, **1 (1)**, 2007, DOI: 10.1117/1.2642347.

Eisenhauer, J.G. Regression through the Origin. *Teaching Statistics*, **25 (3)**, 76–80, 2003, DOI: 10.1111/1467-9639.00136.

Evans, D.S., and M.S. Greer. *Polar Orbiting Environmental Satellite Space Environment Monitor – 2 Instrument Descriptions and Archive Data Documentation*, Natl. Atmos. and Oceanic Admin., Space Environ. Cent, Boulder, Colorado, NOAA Technical Memorandum OAR SEC 93, version 1.4, January 2004, 2000.

Galand, M., and D.S. Evans. *Radiation Damage of the Proton MEPED Detector on POES (TIROS/NOAA) Satellites*, Space Environment Center, Boulder, Colorado, NOAA Technical Memorandum OAR 456-SEC, 2000.

Horne, R.B., S.A. Glauert, N.P. Meredith, D. Boscher, V. Maget, D. Heynderickx, and D.Pitchford. Space weather impacts on satellites and forecasting the Earth's electron radiation belts with SPACECAST. *Space Weather*, **11**, 169–186, 2013, DOI: 10.1002/swe.20023.

McFadden, J.P., D.S. Evans, W.T. Kasprzak, L.H. Brace, D.J. Chornay, et al. *In-Flight Instrument Calibration and Performance Verification*, ESA Publications Division, 277–385, 2007.

Ødegaard, L.-K.G. Recalibration of the MEPED Proton Detectors Onboard NOAA POES Satellites, *Masters degree*, University of Bergen, 2013.

Raben, V.J., D.S. Evans, H.H. Sauer, S.R. Sahm, and M. Huynh. *TIROS/NOAA Satellite Space Environment Monitor Data Archive Documentation: 1995 Update. NOAA Technical Memorandum ERL SEL-86*, National Oceanic and Atmospheric Administration, 1995.

Sandanger, M.I., L.-K.G. Ødegaard, H. Nesse Tyssøy, J. Stadsnes, F. Søraas, K. Oksavik, and K. Aarsnes. In-flight calibration of NOAA POES proton detectors – derivation of the MEPED correction factors. *J. Geophys. Res. [Space Phys.]*, **120**, 9578–9593, 2015, DOI: 10.1002/2015JA021388.

Seale, R., and R.-H. Bushnell. *The TIROS-N/NOAA A-J Space Environment Monitor Subsystem. NOAA Technical Memorandum ERL SEL-75*, Space Environment Laboratory, Boulder, Colorado, 1987.

Yando, K., R.M. Millan, J.C. Green, and D.S. Evans. A Monte Carlo simulation of the NOAA POES medium energy proton and electron detector instrument. *J. Geophys. Res. [Space Phys.]*, **116 (A10231)**, 2011, DOI: 10.1029/2011JA016671.